\documentclass[aps,prb,reprint,nofootinbib]{revtex4-2}

\usepackage{amsmath}
\usepackage{amsfonts}
\usepackage{graphicx}
\usepackage{xcolor}
\usepackage{hyperref}
\usepackage{enumerate}

\newcommand{\unit}[1]{\ensuremath{\mathrm{\ #1}}}

\newcommand{\solar}{\ensuremath{\odot}}

\newcommand{\half}{\ensuremath\frac{1}{2}}
\newcommand{\grad}{\ensuremath\vec{\nabla}}
\newcommand{\td}[2]{\ensuremath\frac{d#1}{d#2}}
\newcommand{\pd}[2]{\ensuremath\frac{\partial #1}{\partial #2}}

\newcounter{probno}
\newcommand{\problem}{\emph{Exercise \arabic{probno}. }\addtocounter{probno}{1}}

\begin{document}

\title{Ultralight Dark Matter: Undergraduate Physics in Modern Cosmology}
%%% Authors will be hidden for blind review
% \author{Removed for Peer Review}
% \affiliation{Removed for Peer Review}
\author{Timothy D.~Wiser}
\affiliation{Truman State University, Kirksville, Missouri, USA}
\email{tdwiser@truman.edu}
\date{\today}
\begin{abstract}
    Ultralight dark matter is a hypothetical class of particle with a number of interesting theoretical and experimental properties, many of which are best understood by direct analogy with or application of undergraduate physics. We present a series of exercises and discussions which may inspire the reader to bring contemporary research on ultralight dark matter into the undergraduate classroom.
\end{abstract}
\maketitle

\section{Background}
    The existence of dark matter (DM)---cold, electrically neutral, gravitating matter outweighing normal matter five to one~\cite{Bertone:2016nfn,ParticleDataGroup:2024cfk}---has a solid empirical justification from observations of galactic rotation curves~\cite{Sofue:2000jx}, galaxy cluster velocity dispersions~\cite{1977MNRAS.179...33W}, gravitational lensing~\cite{Clowe:2006eq}, the cosmic microwave background~\cite{Planck:2018vyg}, and large-scale cosmological structures~\cite{BOSS:2016wmc}.
    In spite of these several lines of evidence, the particle nature of DM is unknown, and has few restrictions: it must be ``cold''---meaning nonrelativistic, $v\ll c$---in order to fall into gravitational potentials and form structures; it must be ``dark''---electrically neutral---to account for its transparancy and inability to radiate away energy and collapse into the disks typical of the normal matter in galaxies~\cite{ParticleDataGroup:2024cfk}.
	(``Clear'' or ``invisible'' matter~\cite{invisible_matter} is perhaps a more apt name than ``dark,'' but the name has stuck.)
	
    For decades, a leading candidate and the target of several ongoing experiments has been the weakly interacting massive particle (WIMP), analogous to a heavier version of the Standard Model neutrino~\cite{Bertone:2016nfn}.
    Particles of this sort are best detected by individual collisions and the resulting energy deposition~\cite{Schumann:2019eaa,LZ:2019sgr}.

\begin{figure*}
	\includegraphics[]{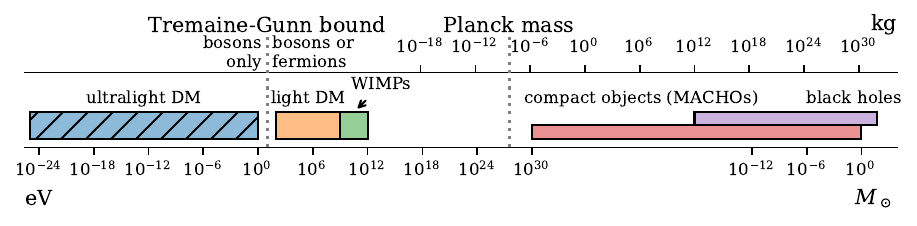}
	\caption{(Color online) Dark matter candidates organized by particle mass (log scale), in units of eV (per $c^2$), kg, and Solar masses ($M_\odot$).
	Only a few of the hundreds of dark matter models are considered here to convey a sense of scale for the parameter space.
	The Planck mass roughly separates elementary particle candidates from ``macroscopic'' candidates, such as black holes~\cite{PBH}, hidden populations of cold brown dwarf stars~\cite{brown_dwarfs}, or other massive compact halo objects (MACHOs)~\cite{MACHO:2000qbb}.
	Among particle DM candidates, the Tremaine--Gunn bound provides a lower limit on the mass of fermions, but not on bosons.
	The Tremaine--Gunn bound and the upper and lower limits on the ultralight DM mass range (hatched region) are explored further in the Exercises.
	The boundary between WIMPs and light DM models represents a difference in detection method and lies around the transition between nuclear recoils and electron recoils in liquid xenon experiments~\cite{LZ:2025igz}.\label{fig:dm_spectrum}}
\end{figure*}

    As experiments continue to rule out more and more possible WIMP models~\cite{LZ:2025igz}, alternative candidates are receiving an increasing share of attention.
	A selection of models are arranged by mass in Fig.~\ref{fig:dm_spectrum}, spanning over 80 orders of magnitude.
    In this Note we will focus on the so-called \emph{ultralight dark matter} scenario~\cite{ferreira_review}, in which individual particle collisions give way to collective interactions with a coherent dark matter field. The upper edge of the ``ultralight'' range is around $1\unit{eV}$ but extends down as far as $10^{-25}\unit{eV}$.
    The ultralight scenario comprises a number of candidate particles, among them axions, axion-like particles (ALPs)~\cite{Graham:2015ouw}, and dark photons~\cite{An:2014twa,Agrawal:2018vin}, which all share some common features.

    Many of the properties of ultralight DM that distinguish it from WIMPs and other DM candidates are best understood by simple arguments that can be adapted from standard undergraduate coursework. For example, the damped harmonic oscillator plays a key mathematical role in understanding the cosmological implications of ultralight DM, and semiclassical arguments from introductory modern physics explain the upper and lower bounds of the ultralight mass range. Experiments to detect ultralight dark matter often center on its interactions with classical electromagnetic fields, rather than individual photons.

    We aim to simultaneously educate the reader about ultralight DM and inspire them to incorporate this cutting-edge physics into their courses via a series of pedagogical sample exercises, arranged by the undergraduate course in which they may best fit.
	Brief solutions are presented in the \hyperref[app]{Appendix}.

\section{Modern Physics}
    Several of the semiclassical concepts typically introduced in modern physics courses have direct relevance to constraining the basic properties of a hypothetical ultralight dark matter particle.
    Note that in the problems below, we use \emph{very} approximate values for the Milky Way galaxy (cf.~the measurements of Ref.~\cite{McMillan:2016jtx}) which make the numbers come out a little bit cleaner: $M\sim 10^{12}M_\solar\sim 2\times 10^{42}\unit{kg}$ and $R\sim 30\unit{kpc}\sim 10^{21}\unit{m}$.\footnote{Some useful constants in mixed astro-particle physics units are $G=4.498\times 10^{-24}\unit{kpc^3\ yr^{-2}\ M_\solar^{-1}}$, $c = 3.066\times 10^{-4}\unit{kpc\ yr^{-1}}$, and $hc=4.018\times 10^{-26}\unit{eV\ kpc}$.}

    \begin{quote} \problem
        Observations suggest a significant total mass of dark matter is gravitationally bound to our own Milky Way galaxy, with a radius of $\sim 30\unit{kpc}$.
        The typical orbital velocity of stars near the edge of the galaxy is $\sim 10^{-3}c$.
        Find the \emph{smallest} mass, in eV, that an ultralight dark matter particle could have, in order for its de Broglie wavelength to ``fit'' in the galaxy.
    \end{quote}
    This is a very simple estimate that explains why experiments searching for ultralight dark matter can assume $m\gtrsim 10^{-24}\unit{eV}$.

    The mass--radius relationship of the galaxy also plays a role in bounding the dark matter mass:
    \begin{quote} \problem
        Model a dark matter particle in a galaxy as a quantum particle in a Newtonian gravitational potential generated by all the other (dark and normal) matter particles with total mass $M$.
        Find the ``gravitational Bohr radius'' $a_0$ of a dark matter particle of mass $m$.
		(In other words---what would the radius of a hydrogen atom's ground state be if its interaction were gravitational instead of electromagnetic?)
        How small can $m$ be for $a_0\le 30\unit{kpc}$ and $M=10^{12}M_\solar$?
    \end{quote}
    Note that both estimates of the minimum mass agree within about an order of magnitude, since the Bohr radius is roughly the minimum possible de Broglie radius of a bound particle consistent with the uncertainty principle, and the orbital velocity follows from the mass and radius. In particular, if $M\sim 10^{12}M_\solar$, then at $R\sim 30\unit{kpc}$ the orbital velocity is roughly $v\sim 10^{-3}c$.
    
    At the bottom edge of the allowable mass range, the orbit of the dark matter is supported by its quantum nature (like an electron in an atom) rather than by its momentum (like a planetary orbit).
    This parameter region may be observationally interesting in its own right, and goes by the name of ``fuzzy dark matter''~\cite{Hui:2016ltb,Eberhardt:2025caq}.

    Is ultralight dark matter a boson or a fermion?
    Fermionic dark matter can be light, but not \emph{ultra}light, thanks to the Pauli exclusion principle:
    \begin{quote} \problem
        If dark matter is made of fermions, then each fermion must be in its own quantum state.
        Estimate the number of quantum states $N$ available for dark matter particles in our galaxy by calculating the total available phase space volume and dividing by the minimum phase space volume set by the uncertainty principle.

        If there is one dark matter particle per state, what mass $m$ must each particle have in order to provide enough gravity for the galaxy ($M=Nm\gtrsim 10^{12}M_\solar$)?
    \end{quote}
    This mass limit is known as the Tremaine--Gunn bound~\cite{tremaine_gunn}.
	It shows, for example, that ordinary neutrinos can't be dark matter, since their masses---though still unmeasured---are known to be less than an electron-volt~\cite{ParticleDataGroup:neutrino_bound}.
    A refined estimate, modeling the fermionic DM as a degenerate Fermi gas, could serve as a nice example in an astrophysics or statistical mechanics course; the rough estimate made above corresponds to a non-interacting Fermi gas, while adding the self-gravity of the DM results in a system analogous to a neutron star~\cite{oppenheimer_volkov}.
    
    Since bosons are not subject to the requirement that there is at most one particle per quantum state, they are allowed to be lighter than fermionic dark matter by many orders of magnitude.
    It is this mass range---roughly $10^{-25}\unit{eV}$ to $1\unit{eV}$--- of bosons that can be considered to be ``ultralight dark matter.''
    Importantly, in this mass range, each quantum state contains many individual bosons.

\section{Classical Mechanics}

    As a bosonic quantum field with high occupation number, an ultralight dark matter candidate is well described not by a collection of classical particles, but as a classical field. (Analogously, when there are many photons per quantum state, we can safely switch to calculating with classical electromagnetism.)

    Furthermore, as dark matter is necessarily cold (nonrelativistic), the spatial variation (momentum) of the field is small compared to its temporal variation (energy).
    So the Lagrangian of the classical field can be further simplified:
    \begin{align}
        L &\equiv \int d^3x\,\mathcal{L} = \int d^3\,x \left(\mathcal{T}(\dot\phi,\grad\phi,\phi) - \mathcal{U}(\phi)\right)\\
        &\simeq (\mathcal{T}(\dot\phi,\phi) - \mathcal{U}(\phi))V \equiv T(\dot\phi,\phi)-U(\phi)
    \end{align}
    ---in other words, the Lagrangian of a classical particle with ``coordinate'' $\phi(t)$.
    Here $V=\int d^3x$ is the total volume of space which is just an overall constant (in the approximation we are making), $\mathcal{T}=\frac{1}{2}\dot\phi^2 - \frac{1}{2}|c\grad\phi|^2 - \frac{1}{2}\omega^2\phi^2$ is the field kinetic energy density, and $\mathcal{U}$ represents any interaction energy (very often negligible for problems involving dark matter!).
    The mass of the DM particle is related to the constant $\omega$ by $mc^2=\hbar \omega$, which can be seen by working out the energy--momentum relation (dispersion relation) for the field.
    
    The analogy between field and classical particle is more mathematical than physical, as this simplified Lagrangian does not describe the motion of the DM particles through space, but rather the evolution of the field's value over time; $\phi(t)$ plays the role of a time-evolving coordinate $x(t)$. \emph{Moving} dark matter particles have field momentum and $\grad\phi\neq 0$, but many times their motion can be ignored unless $v\to c$.

    \begin{quote} \problem
        Given the simplified Lagrangian for a scalar field
        \begin{equation}
            L = \left(\frac{1}{2}\dot\phi^2 - \frac{1}{2}\omega^2\phi^2 - \mathcal{U}(\phi)\right)V,
        \end{equation}
        find (1) the Euler--Lagrange equation of motion for $\phi$ and (2) an expression for the conserved energy density (energy per unit volume of space $V$) in terms of $\phi$ and $\dot \phi$.

        In the absence of any interactions ($\mathcal{U}=0$) what familiar physical system is this?
        Describe, qualitatively and quantitatively, the behavior of $\phi(t)$.
    \end{quote}
	From a high-level point of view, it's not surprising that this Lagrangian reproduces the simple harmonic oscillator; there's an exact mathematical correspondence in quantum field theory between non-interacting particles and quantum harmonic oscillators, which cascades down to non-interacting fields and simple harmonic oscillators in the classical limit.
	What is less obvious---especially to the undergraduate student---is how a harmonic oscillator has anything to do with particles of dark matter in the first place.
	In order to explore the connection, we have to introduce some cosmology; in particular, we will look at how the oscillating field responds to an expanding Universe, to show that it has the same behavior as a collection of ordinary particles at rest.

    In order to account for the Hubble expansion of the Universe, we introduce the scale factor $a(t)$, the size of the Universe's spatial distances relative to today.
	That is, $a(t_0)=1$ today, $a<1$ in the past, and $a>1$ in the future as the Universe continues to expand.
    $a(t)$ has a complicated functional form over all of cosmological history, but can be usefully approximated as a power law or as an exponential during different phases of cosmological history. 
	The Hubble parameter in the distance--redshift relation (Hubble's Law, $v=Hd$) is related to the scale factor by $H=\frac{\dot a}{a}$.
	$H$ is not generally constant unless $a(t)$ is an exponential function of time.

    Neglecting the spatial variations of the field as before, the only effect of Hubble expansion on the Lagrangian is through expanding the volume of space:
    \begin{equation}
        V \to V(t)=(a(t))^3V_0
    \end{equation}
	In this way, some amount of cosmology can be incorporated straightforwardly into a course on classical mechanics, but students will have to simply accept some assumptions about the functional form of $a(t)$.

    \begin{quote} \problem
        Show that in an exponentially expanding Universe with Hubble parameter equal to a constant $H$ (so $a(t)=e^{H(t-t_0)}$), an ultralight scalar field with no interactions undergoes damped harmonic motion.

        Find the conditions on $H$ versus $\omega$ for critical, under-, and over-damping.
    \end{quote}
    The level of damping changes the behavior of the field, and only underdamped oscillations can be interpreted as dark matter.
	The next Exercise explores why:

    \begin{quote} \problem
        The expansion of the Universe breaks the time-translation invariance of the Lagrangian, so Noether's theorem no longer guarantees a conserved energy.
        In fact, something like a work--energy theorem emerges:
        \begin{equation}
			\td{E}{t} = -p\td{V}{t}
		\end{equation}
        with $E=V(t)(\half\dot\phi^2 + \half \omega^2 \phi^2)$ the comoving energy and $p=\half \dot\phi^2-\half \omega^2\phi^2$ the field pressure.

        Prove this relation, and investigate the behavior of the comoving energy in two interesting limits: highly overdamped ($\dot\phi\ll \omega \phi$, $\phi\approx\textrm{const.}$) and highly underdamped ($\phi\sim A(t) \cos(\omega t)$, with the amplitude $A$ varying slowly with time). (Hint: For the underdamped case, average $E$ and $p$ over a whole oscillation of the field, holding $A(t)$ constant over that short time.)
    \end{quote}
	The overdamped case does not behave like dark matter at all---rather, it behaves as a form of energy constant \emph{density} (since $E\sim V\times \textrm{const.}$) even in the face of Hubble expansion.
	The expansion of the Universe ``does work'' on the $\phi$ field by expanding it against a negative pressure, accounting for the increase in total energy.
	In fact, this is exactly the sort of property required of dark \emph{energy}, and it contributes to accelerating the expansion of the Universe.
	(Since a single ultralight field can't be both underdamped and overdamped at the same time, whatever field provides dark matter today can't also constitute dark energy, but some cosmologists speculate on a link between the rapid expansion of early-Universe cosmological inflation and today's dark matter~\cite{inflation_dm}.)
	
    On the other hand, the underdamped field has zero (average) pressure, thus $E$ must be constant with time; as $V(t)$ rises the energy dilutes (slowly decreasing $A$) but the total is constant.
	We can interpret this result by thinking of the ultralight DM as a collection of particles at rest; though their number density decreases like $1/V(t)$, the total number (and thus total mass energy) is conserved.
    In other words, the effect of a large number of ultralight DM particles is modeled by a single classical field undergoing (under)damped harmonic oscillation.

    Many models of ultralight dark matter production in the early Universe depend on the transition from over- to underdamped as the Universe cools and $H$ decreases to set the correct total mass of dark matter~\cite{Marsh:2015xka}.
    The field damping prevents the DM energy density from diluting until a certain point in cosmological history ($H\sim \omega$), allowing for predictions of the dark matter abundance from the mass of the field.
	Since the dark matter abundance today is known, cosmological history puts constraints on the mass and initial conditions of an ultralight DM field.

\section{Electromagnetism}
    Ultralight dark matter candidates are hypothesized to couple with Standard Model particles via one or more interactions.
    For example, most models of axions couple directly to gluons and directly or indirectly to quarks, leptons, and photons.
    In particular, a photon--axion coupling leads to a modification of Maxwell's equations (via a Lagrangian term proportional to $\phi\vec E\cdot \vec B$):
    \begin{align}
        \grad\cdot \vec E &= \rho/\epsilon_0 + g(\grad \phi)\cdot \vec B\\
        \grad\times\vec E &= -\pd{\vec B}{t} \\
        \grad\cdot \vec B &= 0 \\
        \grad\times\vec B &= \mu_0 \vec J + \epsilon_0\mu_0 \pd{\vec E}{t} + g\pd{\phi}{t}\vec B - g(\grad \phi)\times \vec B,%\\
        %g\vec E\cdot \vec B &= \nabla^2 \phi - \frac{1}{c^2}\pd{^2\phi}{t^2} - \omega^2 \phi .
    \end{align}
    where $g$ is a coupling constant representing the unknown strength of the interaction between axions and photons.
	While there is no hard requirement for $g\neq 0$, its value is a consequence of the axion's origin in high-energy physics.
	Two of the earliest and simplest axion models---the KSVZ `hadronic' axion~\cite{ksvz_kim,ksvz_svz} and the DFSZ `leptonic' axion~\cite{dfsz_dfs,dfsz_z}---both predict axion--photon couplings, but with slightly different values.
	Thus, even low-energy measurements using classical electromagnetism can have implications for particle physics models.
    
    \begin{quote}
        \problem
        Suppose a strong, DC magnetic field is produced (e.g.~by a solenoid). What effect does a dark matter axion field $a\propto \cos(\omega t)$ have on the electric and magnetic fields? (Hint: the product of $\pd{\phi}{t}$ and $\vec B$ acts as though it were a current density.)
    \end{quote}
    Existing experiments aim to detect these tiny oscillating fields via highly sensitive magnetometry~\cite{JacksonKimball:2017elr} and/or high-$Q$ resonant cavities~\cite{ADMX:2025vom,Silva-Feaver:2016qhh,Chaudhuri:2014dla}.
    Since the density of DM near Earth is known, at least roughly, based on the Galactic rotation curve, the only unknowns in the experiment are the mass $m$ and coupling $g$, which determine the frequency and amplitude of potential experimental signals.

\section{Discussion}
There are a number of other connections to undergraduate coursework that could be made and we hope the reader may explore some of them: axion-induced electric dipole moments in neutrons, detectable by techniques analogous to NMR~\cite{Graham:2013gfa}; the formation of galaxy-scale Bose--Einstein condensates by ultralight particles~\cite{Guth:2014hsa}; and the production and detection of ultralight particles by interconversion with photons, which can be described as a coupled system of oscillators~\cite{Redondo:2010dp}.

We hope the topics discussed above can help to answer the perennial student question: Why do I have to learn this? A solid foundation in classical and semiclassical physics brings intuition, skills, and analogies to bear on modern research questions, even ones as quantum and cosmological as ultralight dark matter.

\begin{acknowledgments}
	We thank three anonymous referees for their useful feedback which improved the manuscript.
	Many thanks are due also to the author's students who have, from time to time, suffered through unrefined versions of these problems.
\end{acknowledgments}

\appendix

\section*{Appendix: Solutions to Exercises\label{app}}

\begin{enumerate}[\it Ex.~1.]
\item We need $\lambda=\frac{h}{p}=\frac{h}{mv}<R$, so $m>\frac{h}{Rv}$. To use the constants in the given units multiply both sides by $c^2$ to get $mc^2 > \frac{hc}{R(v/c)} = 10^3(4.018\times 10^{-26}\unit{eV\ kpc})(30\unit{kpc})^{-1}\simeq 1.3\times 10^{-24}\unit{eV}$.

\item The usual Bohr radius formula is $a_0=\frac{4\pi\epsilon_0\hbar^2}{me^2}$. To convert to gravitational instead of Coulomb energy, we make the substitution $\frac{e^2}{4\pi\epsilon_0}\to GMm$, giving $a_0=\frac{\hbar^2}{GMm^2}$. In order for $a_0 < R$, we need $m>\frac{\hbar}{\sqrt{GMR}}$ or 
\begin{align*}
	mc^2 &> \frac{(h c)c}{2\pi \sqrt{GMR}}\\&= (2\pi)^{-1}(4.018\times 10^{-26}\unit{eV\ kpc}) \\
	&\quad\times(3.066\times 10^{-4}\unit{kpc/yr})\\
	&\quad\times\Bigl(4.498\times 10^{-24}\unit{kpc^3 yr^{-2} M_\solar^{-1}}\\
	&\quad\times 10^{12}M_\solar \cdot 30\unit{kpc}\Bigr)^{-1/2}\\
	&\simeq 1.6\times 10^{-25}\unit{eV}.
\end{align*}
Note that factoring out an $R$ from the denominator gives $m > \frac{\hbar}{R\sqrt{GM/R}}$, and $\sqrt{GM/R}$ is the velocity of a circular orbit; this limit is really the same as the previous limit except for a factor of $2\pi$ and some rounding of approximate values.

\item The phase space volume is bounded (roughly) by a sphere of radius $R$ in position space `times' a sphere of radius $p$ in momentum space. We should divide this phase space up into tiny hypercubes of volume $(\Delta x)^3(\Delta p)^3\sim \hbar^3$ according to the position--momentum uncertainty principle. Thus the number of states is roughly
\begin{equation*}
	N \simeq \frac{(4\pi/3)^2R^3(mv)^3}{\hbar^3}.
\end{equation*}
(I've left out a factor of 2 for the spin states necessary to have a fermion in the first place---it doesn't change the answer much.) The limit on $m$ is $Nm\gtrsim M,$ or
\begin{align*}
	Nm &= \frac{(4\pi/3)^2R^3m^4v^3}{\hbar^3} \gtrsim M \\
	\implies m &\gtrsim \left(\frac{M\hbar^3}{(4\pi/3)^2R^3v^3}\right)^{1/4} \\
	mc^2 &\gtrsim \left(\frac{Mc^2(h c)^3}{(2\pi)^3(4\pi/3)^2R^3(v/c)^3}\right)^{1/4} \\
	&= \left(\frac{10^{12}M_\solar c^2 (4.018\times 10^{-26}\unit{eV\ kpc})^3}{(128/9)\pi^5(30\unit kpc)^3(10^{-3})^3}\right)^{1/4}\\
	&\simeq 5\unit{eV}
\end{align*}
using $M_\solar c^2 \simeq 1.1\times 10^{66}\unit{eV}.$
We have greatly oversimplified the phase space volume, but thanks to the fourth power of $m$, even fairly large dimensionless factors will not make significant changes to the limit.
For example, including the spin multiplicity for spin-1/2 only shifts the limit down to about $4\unit{eV}$.

\item Applying the Euler--Lagrange equation gives
\begin{equation*}
	\frac{d}{dt}\left(\dot\phi\right) - (-\omega^2\phi - \mathcal{U}'(\phi)) = 0,
\end{equation*}
or more familiarly $\ddot \phi + \omega^2 \phi = -\mathcal{U}'(\phi)$, the harmonic oscillator equation (with an arbitrary external conservative force field). The simple harmonic oscillator results when $\mathcal{U}=0$.

Since this is a time-independent Lagrangian, the conserved energy is the Hamiltonian $H=\dot\phi\frac{\partial L}{\partial \dot\phi}-L = V\left(\frac{1}{2}\dot\phi^2 + \frac{1}{2}\omega^2\phi^2 + \mathcal{U}(\phi)\right)$, and the energy density is functionally the same as the kinetic plus potential energy of the harmonic oscillator, $u=\frac{1}{2}\dot\phi^2 + \frac{1}{2}\omega^2\phi^2 + \mathcal{U}(\phi)$.
Interestingly, the mass of the ultralight particle (part of its kinetic energy) has become equivalent to the harmonic oscillator's elastic potential energy, since $\omega=mc^2/\hbar$.

When $\mathcal{U}=0$ the solutions of $\phi(t)$ are exactly the sinusoidal oscillations with angular frequency $\omega$. The energy density of the field is proportional to the square of the field amplitude.

\item The solution method is the same, but with $V\to a^3V_0$. This results in an additional term in the total time derivative:
\begin{align*}
	\frac{d}{dt}\left(a^3\dot\phi\right) + \omega^2a^3\phi &= 0 \\
    a^3\ddot\phi + 3a^2\dot a \dot\phi + \omega^2 a^3\phi &= 0.
\end{align*}
Dividing through by $a^3$ gives the standard form, $$\ddot\phi + 3H\dot\phi + \omega^2\phi = 0,$$ identical to the damped harmonic oscillator.
Plugging in a solution $\phi=A\exp(kt)$ requires $k=\frac{-3H\pm\sqrt{9H^2-4\omega^2}}{2}$, which is real for $H\ge\frac{2\omega}{3}$. Thus underdamped fields have $\omega > 3H/2$, overdamped fields $\omega < 3H/2$, and critically damped fields $\omega=3H/2$.
The units of $H$ are customarily $\mathrm{(km/s)/Mpc}$, equivalent to frequency, so the dimensions of the result work out.

\item Once can prove this work--energy-like theorem by direct evaluation and substitution of the Euler--Lagrange equation, or from a Hamiltonian-style approach. Starting from $H=\dot\phi\frac{\partial L}{\partial \dot\phi}-L$, $\frac{dH}{dt} = \ddot\phi\frac{\partial L}{\partial\dot\phi} + \dot\phi \frac{d}{dt}\frac{\partial L}{\partial \dot\phi} - \frac{dL}{dt}$.
Substituting the equation of motion and expanding $\frac{dL(\phi,\dot\phi,t)}{dt}$ with the chain rule leads to $\frac{dH}{dt}=-\frac{\partial L}{\partial t} = -(L/V)\frac{dV}{dt}$. Then it is straightforward to identify $H=E$ and $L/V=p$ as defined in the problem.

The overdamped case with $\phi\sim \phi_0 \sim \textrm{const.}$ leads to $E(t)=V(t)(\omega^2\phi_0^2/2)$ and $p=-\omega^2\phi_0^2/2$, constant energy density and constant \emph{negative} pressure. This is not at all what one expects for an expanding gas of massive particles---the energy density should decrease, and the pressure should be positive or nearly zero.

In the underdamped case, $E(t)\simeq V(t)A(t)^2\omega^2$, and $p\simeq A(t)^2\omega^2(\sin^2\omega t - \cos^2\omega t)/2=-A(t)^2\omega^2\cos(2\omega t)/2$---neglecting $\dot A(t)$ in both equations. (This is the slowly-varying limit for the amplitude: $\dot A \ll \omega A$.) The pressure is rapidly oscillating and thus averages to zero; the energy must then be constant (by the work--energy-like theorem) and we must have $A\propto 1/\sqrt{V}\propto a^{-3/2}$. The energy density varies inversely with volume---that is what we would expect for an expanding, interactionless gas; constant total energy spread out over larger volume.

\item We can safely ignore the terms with $\grad \phi$ which are smaller than the time derivatives by a factor of roughly $v/c\sim 10^{-3}$. That leaves just one term, added to the ordinary and displacement current density in the Amp\`ere--Maxwell law. The current density is going to oscillate quickly (at $\omega$) and point in the direction of $\vec B$, along the solenoid's axis of symmetry. Thus we should expect an oscillating magnetic field winding around the solenoid, and a corresponding oscillating Faraday electric field along the solenoid's axis. One could imagine coupling to these oscillating fields via either a dipole antenna along the axis of the solenoid, or a magnetic loop antenna near the solenoid, in a plane through its axis of symmetry.

\end{enumerate}

\bibliographystyle{apsrev4-2}
\nocite{CONTROL}
\bibliography{references}

\end{document}